# Bilirubin lowering effect and safety of a prototype low cost blue light emitting diode (LED) phototherapy device in the treatment of indirect hyperbilirubinemia among healthy term infants in a tertiary government hospital: a pilot study.

Author

**Vanessa Marie V. Calabia, MD**

Co-authors

**Ma. Lucila M. Perez, MD, FPPS**

**Ateneo Innovation Center**

Gregory L. Tangonan, Ph.D.,

Paul M. Cabacungan, MS,

Ivan B. Culaba, MS,

Jeremy E. De Guzman MD., MBA

# ABSTRACT

**Title**: Bilirubin lowering effect and safety of a prototype low cost blue light emitting diode (LED) phototherapy device in the treatment of indirect hyperbilirubinemia among healthy term infants in Ospital ng Makati: a pilot study.

**Background**:
Phototherapy is the most frequently used treatment for the control of neonatal jaundice. The rate of decline of total serum bilirubin level during phototherapy depends on several factors: type of phototherapy, type of light, irradiance level, surface area covered and initial total serum bilirubin level. Light-emitting diodes (LEDs) are power-efficient, low heat-producing light sources that has the potential to deliver high intensity light of narrow wavelength band in the blue-green portion of the visible light spectrum, which overlaps the absorption spectrum of bilirubin resulting in potentially shorter treatment times. These characteristics make LEDs an optimal light source for phototherapy.

**Objective**: This pilot study was done to evaluate the capability of a prototype low cost blue light emitting diode (LED) phototherapy device in lowering bilirubin levels among healthy term infants diagnosed with indirect hyperbilirubinemia.

**Methods**: Experimental study on term infants diagnosed with indirect hyperbilirubinemia in Ospital ng Makati from May 2016 to November 2016 who underwent phototherapy using the low cost blue LED phototherapy prototype.

**Results**: After 24 hours of phototherapy under the prototype LED phototherapy unit, 16% of the total patients completed treatment as they were already classified in the low risk zone, and another 36% of patients completed treatment after 48 hours. The total bilirubin significantly decreased from baseline bilirubin levels after 24 hours by 16.5% ($p = 0.0001$). The mean percentage of change of bilirubin reduced after 48 hours of 29.9% was also significant. The proportion of subjects in the high risk zone during baseline to $24^{th}$ hour went down significantly from 80% to 28% ($p = 0.0003$), while comparing baseline to $48^{th}$ hour, the percentage of high risk zone went down from 80% to 9.5% ($p = 0.0001$). No subjects were reported to have rebound hyperbilirubinemia after discontinuation of phototherapy treatment under the LED prototype. No patient experienced any complication while on phototherapy treatment.

**Conclusion**: The prototype low cost blue light emitting diode (LED) phototherapy was able to lower total serum bilirubin among healthy term infants with indirect hyperbilirubinemia and was safe to use.

**INTRODUCTION**

Jaundice is one of the most common conditions needing medical attention in newborn infants. It is caused by a raised level of bilirubin in the body, a condition known as hyperbilirubinemia. Neonatal jaundice occurs in 25% to 50% of term newborns, and in a larger proportion of preterm newborns, in the first two weeks of life. Most jaundice is a benign transient physiological event in the majority of newborns. Various mechanisms are involved in producing this 'physiological' increase in serum total bilirubin include increased production of bilirubin due to lysis of red blood cells, decreased ability of liver cells to clear bilirubin and increased enterohepatic circulation. Any condition that further increases bilirubin production or alters the transport or metabolism of bilirubin increases the severity of the physiological jaundice.[1] Because of the potential toxicity of bilirubin which can cause permanent neurological disability, it is important to identify newborn infants who might develop severe hyperbilirubinemia.[2]

Phototherapy is the most frequently used treatment when serum bilirubin levels exceed physiological limits. To initiate phototherapy without delay is the most important intervention for infants with severe hyperbilirubinemia. Phototherapy converts bilirubin into water soluble photo-products that can bypass the hepatic conjugating system and be excreted without further metabolism. The clinical response to phototherapy depends on the efficacy of the phototherapy device, as well as the infant's rates of bilirubin production and elimination.[3]

Phototherapy can be delivered using several types of conventional light sources, including daylight, white or blue fluorescent bulbs and filtered halogen bulbs. The efficacy and ability of these light sources to provide intensive phototherapy varies widely and some may be limited because of the inability to keep them close to the infant due to high heat production and unstable broad wavelength light output. In recent years, a new type of light source, light-emitting diodes (LEDs) have been developed and studied as possible light sources for the phototherapy of neonatal jaundice. LEDs are power efficient, portable devices with low heat production, light in weight and have a longer lifetime.[1] Blue LEDs emit a high intensity narrow band of blue light overlapping the peak spectrum of bilirubin breakdown resulting in potentially shorter treatment times. These characteristics of LEDs make them an optimal light sources for a phototherapy device.[4]

**REVIEW OF RELATED LITERATURE**

Jaundice is observed during the first week of life in approximately 60% of term infants and 80% or preterm infants. It is a benign transient physiological event in the majority of newborns but can cause irreversible brain damage and kernicterus in some infants if the serum bilirubin levels are very high.[5] Unconjugated hyperbilirubinemia in which the direct-reacting bilirubin level is less than 15% of serum total bilirubin is the most common form of jaundice seen in newborn infants. Common risk factors for pathological unconjugated hyperbilirubinemia include blood group incompatibility, glucose-6-phosphate dehydrogenase enzyme deficiency, prematurity, instrumental delivery and non-optimal breastfeeding.[1]

Phototherapy is the standard treatment for neonatal jaundice.[2] It converts bilirubin into water soluble photo-products that can bypass the hepatic conjugating system and be excreted without further metabolism.[1] It uses irradiation with blue light to photoisomerize bilirubin into products that can be excreted in bile or urine.[5] The clinical response to phototherapy depends on the efficacy of the phototherapy device, as well as the infant's rates of bilirubin production and elimination. To initiate phototherapy without delay is the most important intervention for infants with severe hyperbilirubinemia.[3] The efficacy of phototherapy is dependent upon wavelength, irradiance, exposed body surface area, distance of the phototherapy, and duration of exposure. Intensive phototherapy is provided by use of high levels of irradiance in the 430 to 490 nm band, usually 30 µW/cm$^2$ per nm or higher delivered to as much of the infant's body surface area as possible.[2]

The commonly used light sources for providing phototherapy include blue fluorescent tubes, compact fluorescent tubes and halogen spotlights. The efficacy and ability of these light sources to provide intensive phototherapy may be limited because of the inability to keep them close to the infant due to high heat production and unstable broad wavelength light output. In recent years, a new type of light source, light-emitting diodes (LEDs) have been developed and studied as possible light sources for the phototherapy of neonatal jaundice. LEDs are power efficient, portable devices with low heat production so that they can be placed very close to the skin of the infants without any apparent untoward effects. They are also durable light sources with an average life span of 20, 000 hours hence with a low cost potential.[1] Blue LEDs emit a high intensity narrow band of blue light overlapping the peak spectrum of bilirubin breakdown and has a higher delivered irradiance resulting in potentially shorter treatment times. These characteristics of LEDs make them an optimal light sources for a phototherapy device.[4]

Vreman et. al. in 1998 studied high intensity light-emitting diodes (LEDs) as possible light sources for the phototherapy of hyperbilirubinemic neonates. They constructed a prototype phototherapy device using 300 blue LEDs and compared it with that of conventional phototherapy devices by measuring the in vitro photodegradation of bilirubin in human serum albumin. The result of the study showed that the prototype device with three focused arrays, each with 100 blue LEDs, generated greater irradiance (> 200 microW.cm-2.nm-1) than any of the conventional devices tested and supported the greatest rate of bilirubin photodegradation: degrading bilirubin by 28% of dark control, followed by blue-green (18% of control), and then white light (14% of control) and least was green light (11% of control). Hence it was concluded in the study that light from LED should be considered a more effective treatment for hyperbilirubinemia than light from other phototherapy devices used.[6]

Kumar P, et. al conducted a systematic review on six randomized controlled trials in 2010 that compared the efficacy of LED phototherapy to conventional phototherapy such as halogen light sources and compact fluorescent light sources in decreasing TSB levels and duration of treatment in neonates with unconjugated hyperbilirubinemia. It was concluded in the study that LED light source phototherapy is as effective in decreasing TSB at rates that are similar to phototherapy with conventional light sources.[1]

In a controlled trial conducted by Mohammadizadeh, et. al in 2012 on preterm infants hospitalized in neonatal intensive care unit who needed conventional phototherapy for uncomplicated indirect hyperbilirubinemia, neonates received phototherapy through devices with LEDs or special blue fluorescent tubes. The 'rate of fall of bilirubin' and 'duration of phototherapy' have been the two main outcomes investigated. Treatment failure defined as the need for exchange blood transfusion or intensive phototherapy, side effects like temperature instability (hypothermia or hyperthermia), skin rash, and burns were also considered. Results of the study showed that LED light source is as effective as fluorescent tubes for the phototherapy of preterm infants with indirect hyperbilirubinemia in terms of decreasing levels of bilirubin. However, the LEDs unit resulted in less frequent hyperthermia, compared to the conventional fluorescent group.[7]

In another randomized controlled study of Colindres, et. al. in 2011, 45 preterm neonates requiring phototherapy were randomized to receive phototherapy using LED-based lights, conventional fluorescent blue lights or conventional halogen lights. The average rate of decrease of bilirubin levels was 0.047-0.037 mg/dl/hour, 0.055-0.056 mg dl/ hour and 0.057-0.045 mg/dl/hour in the groups receiving conventional fluorescent blue light, conventional halogen light and LED phototherapy, respectively. The average duration of phototherapy treatment in the three groups was 108.8-85.9 h, 92.8-38.1 h, 110.4-42.6 h, respectively. In this study, LED phototherapy using simple, low-cost set of lights was as effective of conventional phototherapy.[5]

**SIGNIFICANCE OF THE STUDY**

This study aims to evaluate the safety and bilirubin lowering effect of phototherapy using a prototype low-cost light emitting diode (LED) lights in healthy term infants diagnosed with indirect hyperbilirubinemia. Phototherapy, despite being the recommended standard of treatment for hyperbilirubinemia in newborns, may not be available in developing countries because the devices and replacement bulbs are often too expensive.[8] The results of this study can help provide evidence on whether a prototype low cost blue LED phototherapy device is safe to use and is capable of lowering bilirubin levels, prior to undergoing a clinical trial. This is a preliminary study prior to a randomized controlled trial to ascertain and quantify the capability of the device so that it can be use in resource-challenged institutions especially in developing countries.

**OBJECTIVES**

**General:**

To evaluate the effectiveness of a prototype low cost blue light emitting diode (LED) phototherapy device in lowering bilirubin levels among healthy term infants diagnosed with indirect hyperbilirubinemia.

**Specific:**

1. To measure the rate at which the total serum bilirubin (TSB) changes during the first 24 to 48 hours under low cost blue LED phototherapy device.

2. To determine the average change in bilirubin level using a prototype low cost blue LED phototherapy device.

3. To determine the occurrence of treatment failure among term infants under low cost blue LED phototherapy prototype device.

4. To identify the possible side effects of phototherapy using prototype low cost blue LED phototherapy device.

**METHODOLOGY**

STUDY DESIGN

Experimental study on term infants diagnosed with indirect hyperbilirubinemia admitted at the nursery, OB ward, and pediatric ward of Ospital ng Makati from May 2016 to November 2016. Infants who met the inclusion criteria for the study underwent phototherapy using the low cost blue LED phototherapy prototype. Bilirubin levels were monitored, at most for 48 hours.

INCLUSION/EXCLUSION CRITERIA

Included in the this study were term infants, 37 to 42 completed weeks of gestation with weights appropriate for gestational age, who developed indirect hyperbilirubinemia needing phototherapy within the first 14 days of life. The need for phototherapy was based on the age of the infants in hours and serum total bilirubin that is considered high risk to high intermediate risk indirect hyperilirubinemia which is above the 75$^{th}$ percentile based on the hour specific normogram (Appendix B) but not exceeding a total serum bilirubin level of more than 25mg/dl. Infants included had normal physical examination findings and normal laboratory test results such as complete blood count, reticulocyte count, peripheral blood smear and Coomb's test. Excluded from the study were infants with onset of jaundice during the first 24 hours after birth; with direct bilirubin levels more than 20% of total bilirubin; with evidence of hemolysis (positive direct Coombs test); diagnosed with neonatal sepsis; who needed exchange transfusion at the time of enrollment; and those with congenital malformations.

SAMPLE SIZE

Subjects were recruited consecutively as they were diagnosed and deemed eligible based on inclusion/exclusion criteria. A total of 27 patients were recruited for this pilot study. The sample size was computed using the 24 hour decrease in bilirubin of the first 8 subjects that were enrolled in this study and the following conditions: power = 80%, alpha = 0.05, effect size = 1, standard deviation (SD) = 1.854.

INTERVENTION

Baseline characteristics such as sex, weight, age of gestation at birth, age at the start of phototherapy, feeding history, maternal blood type, patient's blood type, cause of hyperbilirubinemia and baseline total serum bilirubin levels of qualified neonates were recorded. The study was explained to the parents prior to enrollment and an informed consent was obtained. All infants enrolled in the study received phototherapy using the blue light emitting diode (LED) phototherapy.

The prototype low cost LED phototherapy unit was custom made in collaboration with Ateneo Innovation Center, a non-profit research organization in Ateneo de Manila University under the School of Science and Engineering. The following specifications were used as guide in making of the prototype: light source using blue light emitting diode (LED) giving off light in the wavelength spectrum of 460 to 490-nm (blue-green light region) with peak spectral irradiance of $\geq$30uW/cm2/nm at a recommended treatment distance of 30 cm above the patient.

The prototype low cost LED phototherapy was made using 20 pieces blue LED lamps, connected in parallel, for a total power rating of 60 W as shown in Figure 1. A Spectro-Vis Plus (Vernier) was used to determine the spectrum of the LED lamps. The range of the emission spectrum was found to be 462.1-476.4 nm (Figure 2). This range falls within the most effective range (460-490 nm) for phototherapy.[9]

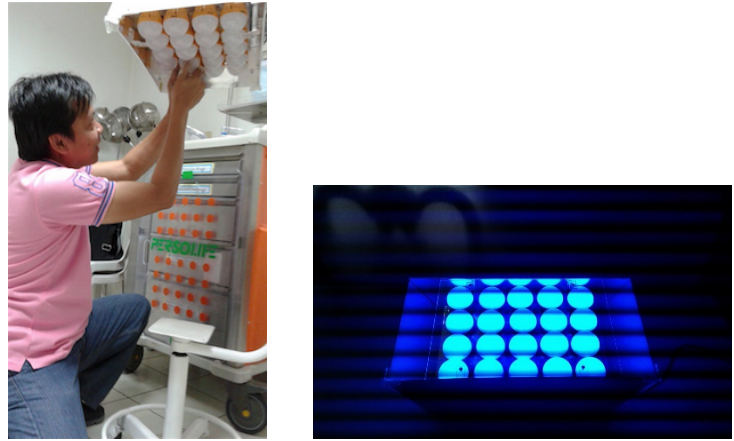

Figure 1. Assembly of the 20 pieces of blue LED lamps.

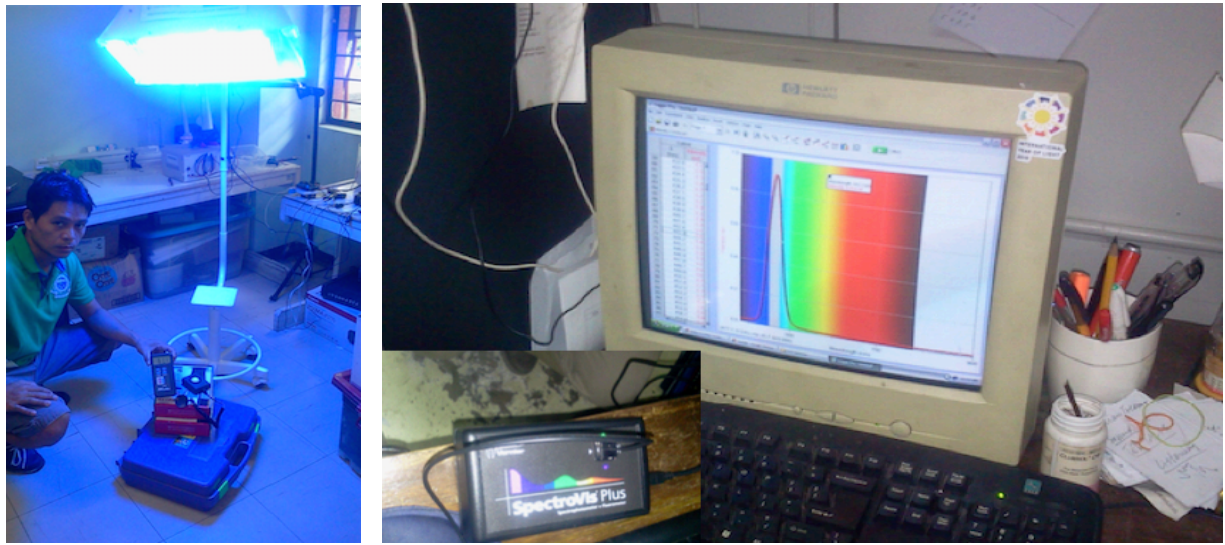

Figure 2. The Spectro-Vis Plus used in determining the emission spectrum

The intensity of the LED lamp was 12W/m$^2$, measured using a light intensity meter. The calculated value of the irradiance was 84 µW/cm$^2$/nm. This value is almost three times higher than the minimum recommended irradiance (30 at µW/cm$^2$/nm) required for intensive phototherapy at 1-m distance.[9] Figure 3 shows the measurement of the intensity of the LED lamp. The calculation of the irradiance is shown below.

Irradiance = measured intensity (microwatt/ square cm) / range of wavelength

Irradiance = $\{(12 \text{ W/ m}^2) * (10^6 \text{ μW / watt}) * [(1 \text{ m}^2/10^4 \text{cm}^2)]\}$
/ {476.4 nm– 462.1 nm}
= 84 μW /cm$^2$/nm

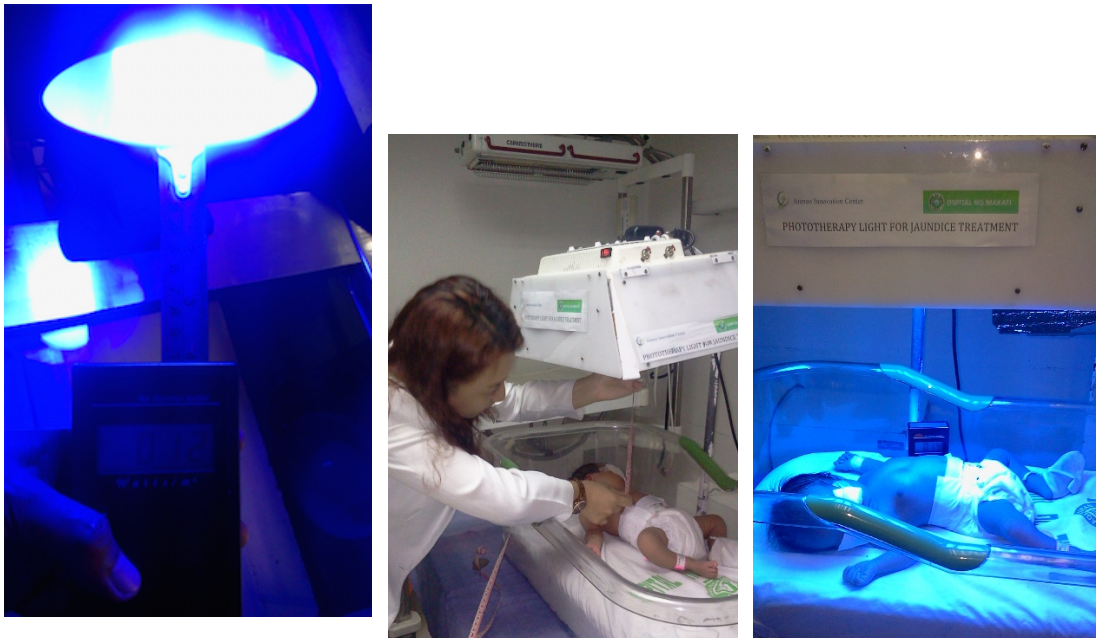

Figure 3. Measurement of the intensity of the LED lamp

The total power consumption of this prototype is 75 watts which is less than one fifth of the commercial phototherapy light (about 400 watts) that uses fluorescent tubes. This prototype unit can also be powered by photovoltaic cells (solar panels). For a 100-watt solar panel at four-hours of sunlight, it can operate for 5 hours with all light bulbs continuously switched on. It is designed in such a way that it can also operate at half power and twice the time.

An internal continuous test-run was conducted with all the lamps turned on for 7-hours, 24-hours, and 72-hours. There was no notable degradation of the quality of the lamp.

The unit was designed in such a way that the whole light casing can be tilted from side to side, the metallic pipe stand height can be adjusted, the whole unit can be rolled through its wheels for mobility/easy handling. Figures 5 and 6 shows the complete phototherapy equipment. The total cost for the development of the prototype equipment was approximately PhP 20,000.00

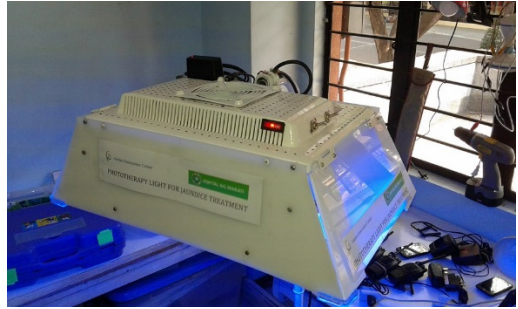

Figure 5. The casing of the light source

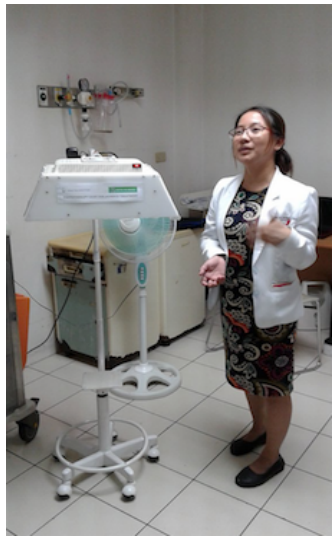 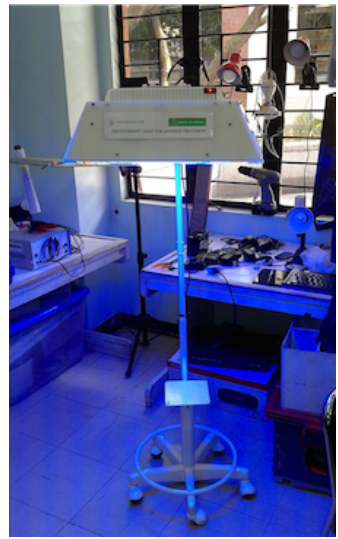

Figure 6. The complete phototherapy equipment

Detailed description of the development of the prototype LED phototherapy unit, including the testing of the equipment and the safety precautions built into the phototherapy device prior its use for the study is further discussed in Appendix F.

Infants were exposed, completely unclothed with their eyes and genital regions properly covered, to continuous phototherapy for 24 hours only interrupted during feeding, cleaning, blood sampling, and turning to sides every 2 hours. The interruptions in phototherapy were monitored and mothers were instructed to write down the number and duration of interruptions such as breastfeeding, cleaning and blood extractions in a data collection form that was provided. Those who developed comorbidities including hospital-acquired infections while enrolled in the study were withdrawn from the study. Phototherapy using the prototype blue LED phototherapy unit was done for 24 to 48 hours. Infants, who after the first 24 hours of phototherapy showed an increase in total serum bilirubin were shifted to conventional phototherapy. Subjects with no change in total serum bilirubin level after 24 hours continued to receive treatment using the prototype blue LED phototherapy for another 24 hours. After 48 hours of using the prototype blue LED phototherapy unit, subjects who required further phototherapy treatment were shifted to conventional phototherapy.

Participants enrolled in the study were monitored daily by the nurse-in-charge, including daily weight, vital signs every 4 hours, urine output and occurrence of possible complications such as rashes, burns, diarrhea, and hypothermia or hyperthermia every shift. Occurrence of such complications were managed accordingly: hydration for occurrence of diarrhea and hyperthermia, application of therapeutic ointments for rashes or burns and possible referral to other services such as Dermatology and Surgery as needed.

Total serum bilirubin levels were obtained through blood extraction using peripheral venipuncture at baseline then every 24 hours thereafter upon exposure to phototherapy. Serum bilirubin levels was analyzed in the laboratory of Ospital ng Makati using clinical chemistry analyzer Architect C4000 model by Abbott.

When the infants' total serum bilirubin level fell to the low intermediate risk level (below the 75$^{th}$ percentile of the nomogram or TSB < 14 mg/dl) they were observed for another 24 hours after discontinuation of phototherapy prior to discharge to ensure that there was no rebound hyperbilirubinemia.

If subjects were noted to be again jaundiced 24 hours after discontinuation of phototherapy, total serum bilirubin levels were taken to determine rebound in the bilirubin level. Neonates with rebound bilirubin values greater than the 75th percentile or equal to the baseline total serum bilirubin levels on admission were again exposed to phototherapy but this time under conventional fluorescent phototherapy.

OUTCOMES

Primary outcome results include the average rate of change of the total serum bilirubin level (mg/dl per hour) during the first 24 hours and on the second day of phototherapy treatment. The overall rate of decrease of bilirubin was calculated as initial bilirubin concentration minus final bilirubin concentration over total treatment time (24 to 48 hours). Secondary outcomes determine the occurrence of treatment failure which include increase in TSB levels after 24 hours or no change in TSB level after 48 hours of exposure under prototype low cost blue LED phototherapy and occurrence of possible side effects of phototherapy such as skin rash, burns, diarrhea dehydration, hypothermia and hyperthermia.

ANALYSIS

Descriptive statistics such as mean and standard deviation (SD) were used to present the profile and total serum bilirubin of patients. The mean bilirubin levels and average rate of decline of the total serum bilirubin level with $\pm$ 1 standard deviation were analyzed. The proportion of neonates who developed treatment failure and complications to phototherapy intervention were determined. In comparing the baseline bilirubin against the 24th and 48th hour bilirubin, paired t-test was used. Chi square and fisher test were used in comparing the risk zone level between and baseline vs 24th and 48th hour. Test of significance was at 5% level.

Medcalc statistical software was used to carry out the computations.

**ETHICAL CONSIDERATION**

A voluntary informed consent (Appendix A) was obtained from the parents or legal guardian of the subjects prior to their inclusion in the study. Confidentiality of patients' information was maintained at all times through data anonymity. Only the investigator and those involved in data processing were given access to these records. The study protocol was approved by the hospital research ethics committee.

Subjects were monitored regularly, including daily weight, vital signs every 4 hours, urine output every shift, and occurrence of possible complications such as burns, temperature instability, dehydration, skin rashes and loose stools. The prototype LED phototherapy unit was regularly checked and calibrated once a month.

**RESULTS**

**Profile of Patients**

Twenty seven patients were enrolled in this study however only 25 subjects were analyzed. The 2 subjects were withdrawn from the study due to development of symptoms other than jaundice such as fair suck, fair activity, hypoglycemia and fever, which were attributed with neonatal sepsis. These symptoms were noted after they have completed 48$^{th}$ hour of phototherapy under the LED prototype.

Of the 25 infants, 13 were males and 12 were females. Analyzed infants have an average age of gestation of almost 39 weeks (38.5 ± 1.0), with mean gestational age upon admission of 5 days (5.4 ± 2.1), with mean weight of 3 kg (3.1 ± 0.4). Subjects were mostly blood type O (36%) while the least type was blood type AB. Same pattern can be observed in terms of maternal blood type, where Type O is the most common. In term of diagnosis, 52% were with breastfeeding jaundice, 36% with ABO incompatibility, mostly OA setup and 12% with breastmilk jaundice. All patients were exclusively breastfeeding. Seventeen subjects (68%) were supplemented with intravenous fluid. The mean total serum bilirubin (TSB) of subjects upon admission was 19 mg/dl (19.72 ± 2.93) corresponding above the 95$^{th}$ percentile track (high risk zone) on the hour specific bilirubin nomogram.

| Table 1. Demographic and Clinical Data of Patients (*N*=25) | |
|---|---|
| **Age of Gestation (weeks), mean ± sd** | 38.5 ± 1.0 |
| **Age of Baby (days), mean ± sd** | 5.4 ± 2.1 |
| **Gender, n, %** | |
| Female | 12 (48.0) |
| Male | 13 (52.0) |
| **Weight (kg), mean ± sd** | 3.1 ± 0.4 |
| **Blood Type, n, %** | |
| A | 7 (28.0) |
| B | 8 (32.0) |
| O | 9 (36.0) |
| AB | 1 (4.0) |
| **Maternal Blood Type, n, %** | |
| A | 3 (12.0) |
| B | 3 (12.0) |
| O | 18 (72.0) |
| AB | 1 (4.0) |
| **Diagnosis, n, %** | |
| Breastfeeding jaundice | 13 (52.0) |
| Breastmilk jaundice | 3 (12.0) |
| ABO incompatibility | 9 (36.0) |
|     OA set-up | 6 (24.0) |
|     OB set-up | 3 (12.0) |
| **Breastfeeding, n, %** | 25 (100.0) |
| **With intravenous fluid, n, %** | 17 (68.0) |
| **Without intravenous fluid, n, %** | 8 (32.0) |
| **Baseline TSB (mg/dl), mean ± sd** | 19.72 ± 2.93 |

**Outcome of Treatment**

Table 2 show that after 24 hours of phototherapy under the prototype LED phototherapy unit, 16% of the total patients already completed treatment as they were already classified in the low risk zone, while another 36% of the patients completed treatment after 48 hours. Overall, after 48 hours, 52% of infants completed phototherapy treatment. However, 48% of subjects still needed further treatment after 48 hours of exposure under the LED phototherapy light, most of these patients had baseline total serum bilirubin > 20 mg/dl. Subjects who completed phototherapy after the $24^{th}$ and $48^{th}$ hours were all discharged with good general condition. No subjects were reported to have rebound hyperbilirubinemia after discontinuation of phototherapy treatment under the LED prototype.

| Table 2. Overall outcome of treatment | |
|---|---|
| | Number of subjects, n (%) |
| Completed treatment after 24 hours | 4 (16.0) |
| Completed treatment after 48 hours | 9 (36.0) |
| Continued treatment after 48 hours | 12 (48.0) |

**Comparison on the Total Serum Bilirubin**

Considering all patients whether they have received intravenous fluid (IVF) or not, the mean decrease in bilirubin on the $24^{th}$ hour was 3.25 ± 2.38, with percentage change of 16.5% ($p = 0.0001$). This means that the amount of total bilirubin reduced after 24 hours of treatment under LED prototype was already significantly lower than the baseline. Likewise, the mean percentage of decrease of bilirubin after 48 hours of 29.9% was also significant when compared to baseline ($p = 0.0001$). Moreover, the percentage of change from $24^{th}$ to $48^{th}$ hour of 18.9% was also significant ($p = 0.0001$).

| Table 3. Change in total serum bilirubin (TSB) from baseline vs 24th and 48th hour | | | | |
|---|---|---|---|---|
| TSB | Mean ± SD | p value | Mean Difference ±SD | % Change (95% CI) |
| Baseline (n=25) | 19.72 ± 2.93 | 0.0001 | 3.25 ± 2.38 | 16.5% (11.5% to 21.5%) |
| 24th Hour (n=25) | 16.47 ± 3.10 | | | |
| Baseline (n=21) | 19.72 ± 2.93 | 0.0001 | 5.89 ± 2.71 | 29.9% (23.6% to 36.2%) |
| 48th Hour (n=21) | 14.12 ± 2.98 | | | |
| 24th Hour (n=21) | 17.42 ± 2.32 | 0.0001 | 3.30 ± 1.72 | 18.9% (14.5% to 23.5%) |
| 48th Hour (n=21) | 14.12 ± 2.98 | | | |

Among 17 patients supplemented with intravenous fluid while on phototherapy, the percentage of decrease of 18.4% (p = 0.0001) from baseline to 24th hour was significant. Likewise, the 30.4% decrease in total bilirubin reduced after the 48th hour was also significant. Lastly, the percentage of change from 24th hour to 48th hour was also significant (17.5%).

| Table 4. Change in TSB from baseline vs 24th and 48th hour among patients with IVF | | | | |
|---|---|---|---|---|
| With IVF | Mean± SD | p value | Mean Difference±SD | % Change (95% CI) |
| Baseline (n=17) | 21.15 ± 2.39 | 0.0001 | 3.89 ± 2.42 | 18.4% (12.5% to 24.3%) |
| 24th Hour (n=17) | 17.26 ± 3.26 | | | |
| Baseline (n=14) | 21.15 ± 2.39 | 0.0001 | 6.44 ± 2.91 | 30.4% (22.5% to 38.4%) |
| 48th Hour (n=14) | 15.18 ± 2.82 | | | |
| 24th Hour (n=21) | 18.4 ± 2.24 | 0.0001 | 3.22 ± 1.65 | 17.5% (12.3% to 22.7%) |
| 48th Hour (n=14) | 15.18 ± 2.82 | | | |

For the 8 patients without IVF, the mean decrease in bilirubin was 1.88 ± 1.71 (p = 0.0086) which indicates that the percentage of change of 11.3% from baseline to 24th hour was significant. Likewise, the 28.6% total bilirubin reduced after 48 hours was also significant. And lastly, the percentage of change from 24th hour to 48th hour was also significant (17.6%).

| Table 5. Change in TSB from baseline vs 24th and 48th hour among patients without IVF ||||
| Without IVF | Mean ± SD | p value | Mean Difference ± SD | % Change (95% CI) |
|---|---|---|---|---|
| Baseline (n=8) | 16.66 ± 0.84 | 0.0086 | 1.88 ± 1.71 | 11.3% (2.7% to 19.8%) |
| 24th Hour (n=8) | 14.79 ± 2.01 | | | |
| Baseline (n=7) | 16.80 ± 0.80 | 0.0004 | 4.8 ± 2.02 | 28.6% (17.4% to 39.7%) |
| 48th Hour (n=7) | 12.0 ± 2.13 | | | |
| 24th Hour (n=7) | 15.47 ± 0.59 | 0.0018 | 3.47 ± 1.97 | 17.6% (8.4% to 26.8%) |
| 48th Hour (n=7) | 12.0 ± 2.13 | | | |

**Outcome in terms of Total Bilirubin Risk Zone Classification**

Considering the risk zone classification, the p value of 0.0001 indicates that the improvement is significant. The proportion of subjects in the high risk zone during baseline to 24th hour went down significantly from 80% to only 28%, while increasing also the low intermediate risk and low risk zone classification. Likewise, comparing baseline to 48th hour, the percentage of high risk zone went down significantly from 80.0% to just 9.5%, while increasing low intermediate risk and low risk zone. This suggest that the condition of patients significantly improved after the 24th and 48th hours of treatment under the LED prototype.

| Table 6. Change in total bilirubin risk zone classification ||||
| Risk Zone | Baseline | 24th Hour | 48th hour |
|---|---|---|---|
| High | 20 (80.0) | 7 (28.0) | 2 (9.5) |
| High Intermediate | 5 (20.0) | 10 (4) | 6 (28.6) |
| Low Intermediate | 0 (0.0) | 4 (16.0) | 5 (23.8) |
| Low | 0 (0.0) | 4 (16.0) | 8 (38.1) |

Baseline vs 24th hour p = 0.0003
Baseline vs 48th hour p = 0.0001

**Interruption to phototherapy**

In terms of feeding interruption, mean duration of breastfeeding was approximately 38 minutes every 3 hours in 24 hours while in terms of cleaning, the average cleaning time was 7.5 minutes 4x a day.  Interruption to phototherapy such blood extraction is identical for all patients which is 5 minutes once a day.  Turning was done every 2 hours for all patients.

| Table 7. Interruption to phototherapy | | |
|---|---|---|
| **Interruption** | | **Duration** |
| Breastfeeding | 37.5 ± 5.5 | Every 3 Hours |
| Cleansing | 7.5 ± 0.0 | minutes 4 times a day |
| Blood Extraction | 5 | minutes a day |
| Turning | 2 | Hours |

**Complications**

No patient experienced complication while on phototherapy treatment.  None of the subjects developed hypothermia or hyperthermia while on phototherapy, lowest temperature recorded among the subjects was 36.7C, while the highest temperature was 37.5C.

All subjects had an adequate urine output while on phototherapy treatment under the LED prototype.  Results showed that the mean urine output was 4.37cc/kg/hr (SD ± 0.80).  The highest urine output recorded was 6.5 cc/kg/hr while lowest was 2.9 cc/kg/hr.  No complications such as rashes, blisters nor watery stools was noted.

**DISCUSSION**

Phototherapy has been the mainstay of management of unconjugated hyperbilirubinemia in newborns. Its efficacy is dependent upon irradiance (light intensity), the quality of the light (optimal in the blue-green, 400-550 nm), the exposed body surface area, and the duration of exposure. When initiated early, phototherapy can reduce serum bilirubin levels to within clinically acceptable levels. A new type of light source, light emitting diodes (LEDs), for phototherapy, have a narrow spectral band of high intensity monochromatic light that overlaps the absorption spectrum of bilirubin.[10]

Standard phototherapy units deliver a spectral irradiance of 8 to 10 µW/cm2 per nm. Intensive phototherapy delivers high levels of irradiance in the 430 to 490-nm band (> 30 µW/cm2 per nm) to as much of the infant's surface area as possible. Special blue tubes provide light predominantly in the blue-green spectrum and are highly effective. At wavelengths in the blue-green spectrum, light penetrates the skin well and is absorbed maximally by bilirubin. A 6% to 20% reduction can be expected in the first 18 to 24 hours of conventional phototherapy. Intensive phototherapy can result in a decline of at least 2 to 3 mg/dL (34–51 µmol/L) within 4 to 6 hours. A decrease in total serum bilirubin can be noted as early as 2 hours after initiation of treatment. In infants aged 35 weeks of gestation and older, 24 hours of intensive phototherapy can result in a 30% to 40% decrease in total serum bilirubin levels.[1]

In an unpublished study by Marasigan KL (2014), it was reported that there were about 2.1% and 2.6% of neonates who were admitted due to hyperbiliribunemia in years 2012 and 2013, respectively in Ospital ng Makati. Indirect (Unconjugated) hyperbilirubinemia was found to be more common (79.5%) than conjugated hyperbilirubinemia (20.5%) and out of the 91 recorded cases of indirect hyperbilirubinemia, admitted patients required phototherapy with an average duration of 4.64 days and an average rate of decrease in bilirubin was 0.91 mg/dl per day using conventional phototherapy that uses blue and white fluorescent bulbs.[11] The conventional fluorescent phototherapy in Ospital ng Makati is being administered with an Olidef CZ Medphoto 6 Phototherapy Unit which has six fluorescent lamps in four 20 watts day light and two blue fluorescent with a lamp life-time of 2,000 hours, emitting a light in the main radiation spectrum in the range between 400nm and 550nm, giving an irradiance of 15uW/cm2/nm at 30 cm distance from lamp box.

In this pilot study, phototherapy treatment was given using a low cost blue LED prototype device that was developed aimed to provide intensive phototherapy that generates greater irradiance than the conventional device. Phototherapy using low-cost LED lights was found to be effective in decreasing serum bilirubin levels. As shown in the results, 52% of infants completed phototherapy treatment after 48 hours (Table 1). Although 48% of subjects needed further phototherapy treatment, significant improvement in terms of total bilirubin risk zone classification was already achieved after the 24$^{th}$ and 48$^{th}$ hours of treatment under the LED prototype (Table 4). Furthermore, none of the neonates from the study who completed phototherapy treatment required restarting the phototherapy after 24 hours of having finished the treatment.

Complications associated with phototherapy include loose stools, overheating, dehydration due to increased insensible water loss or diarrhea, hyperthermia, skin rashes and burns. Maintaining adequate hydration and urine output during phototherapy is important since urinary excretion of lumirubin is the principal mechanism by which phototherapy reduces total bilirubin. The American Academy of Pediatrics (AAP) states that intravenous hydration of infants receiving phototherapy is not routinely indicated. However some infants with high bilirubin levels, who are also mildly dehydrated, may need supplemental fluid intake to correct their dehydration. When severe jaundice appears in breastfed infants, the recommendation of AAP includes supplementing breast-feeding with formula in an attempt to increase the caloric intake and decrease the enterohepatic circulation. Continuing fluid supplementation may reduce bilirubin by decreasing entrohepatic cycle of bilirubin and increasing urine output.[2]

In this pilot study, results showed that there was significant decrease in total serum bilirubin level in subjects among both those with and without intravenous fluid supplementation. Although the mean difference and % change of TSB from baseline vs 24[th] and 48[th] hour seem higher in subjects with IVF, 3.89 ± 2.42 mg/dl (18.4%) vs 1.88 ± 1.71 mg/dl (11.3%), baseline TSB of these subjects were also higher to begin with. The baseline mean TSB levels on enrollment in supplemented and non-supplemented groups with IVF were 21.15 ± 2.39 mg/dl and 16.66 ± 0.84 mg/dl, respectively. Also, we found that supplementation fluid had no effect on the duration of phototherapy. Most of the subjects with IVF fluid needed to continue phototherapy beyond 48 hours given their higher initial TSB.

In a randomized controlled study by H. Al-Masri in 2012, administration of intravenous fluid supplementation during phototherapy on healthy full-term jaundiced infants, showed no benefit on reducing the serum bilirubin during phototherapy, and they had similar rates of decrease in TSB levels comparing with newborns given only oral feeding.[12]

In another controlled study by Z. Easa in 2010 that include fluid supplementation in healthy term neonates with non-hemolytic hyperbilirubinemia, intravenous fluid of 1/5 normal saline and 5% dextrose was given to 32 neonates while other 32 neonates had no other additional fluid. There was no significant difference between the two groups in the rate of decreasing TSB or duration of phototherapy hence concluding that administration of extra intravenous fluid in jaundiced healthy, term, breastfed neonates have no beneficial effect on the rate of serum bilirubin reduction during conventional phototherapy.[13]

Side effects like hypothermia and hyperthermia were not found among the subjects in this study. This may be because the enrolled neonates were treated in temperature controlled environments with regular monitoring of body temperature. Since LED does not produce much heat, hypothermia may be a problem hence closer monitoring and external heat source may be required. In a recent report by M. Yurdakok, it was stated that LED phototherapy with high irradiances (60-120 µW/cm2/nm) significantly increases body temperature in hyperbilirubinemic newborns compared to infants who received conventional phototherapy with fluorescent lamps (10-15 µW/cm2/nm) or LED phototherapy (26-60 µW/cm2/nm). Thus the increase in body temperature is also reported to be a function of increase of irradiance rather than the type of the light source. Hyperthermia might be related to release of pyrogenic cytokines, although effects of light with different wave-lengths and irradiances on serum cytokine levels are not known.[14] The LED prototype in this study that emits high irradiance of 84 µW /cm$^2$/nm did not cause hyperthermia among the subjects. Other complications such as loose stools, overheating, dehydration and others (burns and rashes) were also not observed.

**CONCLUSION / RECOMMENDATION**

The prototype low cost blue light emitting diode (LED) phototherapy that was developed was able to lower total serum bilirubin among healthy term infants with indirect hyperbilirubinemia and was safe to use. However, since this is only a pilot study with limited number of subjects, further studies using the prototype LED phototherapy and comparing it with the conventional fluorescent phototherapy in the treatment of indirect hyperbilirubinemia are needed to test its efficacy. Evaluation of the efficacy of the prototype LED phototherapy in jaundiced neonates with sepsis, hemolytic versus non-hemolytic jaundice and in term versus preterm neonates must also be done. The prototype that was developed has the potential to enable phototherapy to be safely and reliably delivered in low-resource settings.

**REFERENCES**

1. Kumar P, Chawla D, Deorari A. Light-emitting diode phototherapy for unconjugated hyperbilirubinaemia in neonates. Cochrane Database of Systematic Reviews 2011, Issue 12. Art. No.: CD007969.

2. American Academy of Pediatrics Subcommittee on Hyperbilirubinemia. Management of hyperbilirubinemia in the newborn infant 35 or more weeks of gestation. Pediatrics 2004; 114:e297–316.

3. Managing Newborn Hyperbilirubinemia and Preventing Kernicterus Progeny, Volume XX1X, No 1, June 2013 available in http://www.idph.state.ia.us/hpcdp/common/pdf/perinatal_newsletters/progeny_june2013.pdf

4. Chang YS, et. al. In vitro and in vivo Efficacy of New Blue Light Emitting Diode Phototherapy Compared to Conventional Halogen Quartz Phototherapy for Neonatal Jaundice. J Korean Med Sci 2005; 20: 61-4

5. Colindres JV, et. al. Prospective Randomized Controlled Study Comparing Low-Cost LED and Conventional Phototherapy for Treatment of Neonatal Hyperbilirubinemia. Journal of Tropical Pediatrics, 2011

6. Vreman HJ, et. al. A light emitting diode (LED) device for phototherapy of jaundice newborns: in vitro efficacy. Pediatr Res 1997;41:185A.

7. Mohammadizadeh M, Eliadarani FK, Badiei Z. Is the light-emitting diode a better light source than fluorescent tube for phototherapy of neonatal jaundice in preterm infants?. Adv Biomed Res 2012;1:51.

8. Malkin R. and Anand V., A novel phototherapy device: the design community approach for the developing world. IEEE Eng Med Biol Mag. 2010 Mar-Apr;29(2):37-43.

9. Bhutani VK and The Committee on Fetus and Newborn. Phototherapy to prevent severe neonatal hyperbilirubinemia in the newborn infant 35 or more weeks of gestation. American Academy of Pediatrics. Pediatrics 2011; 128:e1046–1052.

10. Rosen H, et. al., Use of a Light Emitting Diode (LED) Array for Bilirubin Phototransformation. Conf Proc IEEE Eng Med Biol Soc. 2005;7:7266-8.

11. Marasigan KL and Perez MA. Prevalence and profile of neonates who developed hyperbilirubinemia admitted in a tertiary government hospital. Ospital ng Makati. Unpublished. November 2014.

12. Al-Masri HA. In healthy baby with severe jaundice do we need to give fluid supplementation during phototherapy? Sudan Med J 2012 December;48(3)

13. Easa ZO. Effect of intravenous fluid supplementation on serum bilirubin level during conventional phototherapy of term infants with severe hyperbilirubinemia QMJ Vol.9 No.15, July 2010

14. Yurdakök M. Phototherapy in the newborn: what's new? Journal of Pediatric and Neonatal Individualized Medicine 2015;4(2):e040255

**APPENDICES**

Appendix A.

## Consent Form

Title: **Bilirubin lowering effect and safety of a prototype low cost blue light emitting diode (LED) phototherapy device in the treatment of indirect hyperbilirubinemia among healthy term infants in Ospital ng Makati: a pilot study.**

Ang katibayang ito ay nagbibigay ng impormasyon ukol sa pagsusuring medikal. Sa oras na maunawaan ko ang nilalaman nito, ako ay hihingan ng lagda kung pahihintulutan kong mailahok and aking anak.  Nauunawan ko na gagawa ng pagsusuri ukol sa mabuting idudulot ng paggamit ng LED phototherapy unit sa paglunas ng paninilaw ng aking anak.

Ipinaliwanag sa akin na ang mga ilalahok sa pag-aaral na ito ay yaong mga sanggol na ipinanganak na husto sa buwan at walang ibang karamdaman maliban sa pagkakaroon ng mataas ng lebel ng bilirubin na nagiging sanhi ng kanilang paninilaw.  Ipinaliwanag sa akin na ang LED phototherapy na gagamitin ay nabuo sa tulong ng Ateneo Innovation Center, na ito ay naiiba sa karaniwang ginagamit ng ospital na conventional phototherapy gamit ang fluorescent light.  Naintindihan ko na layunin ng pag-aaral na ito na subukin ang kakayahan ng LED phototherapy upang mapababa ang lebel ng bilirubin ng aking anak.  Aking nauunawaan na sa pamamagitan ng paglahok ng aking anak sa pagsusuri, ito ay maaring makadagdag sa makasiyensyang pamamaraan sa pagpapabilis ng pagbaba ng paninilaw ng aking anak.

Naipaliwanag po sa akin na kukuhanan ng dugo ang aking anak bago magsimula ang pagpapailaw, tuwing 24 oras habang nagpapailaw hanggang sa bumaba ang lebel ng paninilaw at 24 oras matapos itigil ang pagpapailaw sa kanya upang malaman kung nanatiling mababa ang lebel ng bilirubin sa dugo na nagdudulot ng paninilaw at kung sakaling tumaas muli ang lebel ng paninilaw ay kakailanganin syang pailawan muli.  Naipaliwanag din po sa akin at aking naintihdihan na kakailanganing pailawan ang aking anak sa ilalim ng LED phototherapy ng tuloy-tuloy at magagambala lamang ito sa tuwing kailangan nyang dumede, linisan at kuhanan ng dugo.  Ipinaliwanag din na kapag bumaba na ang lebel ng bilirubin ng aking anak matapos ang 48 hours na pagpapailaw sa ilalim ng LED phototherapy ay isasailalim na siya sa pagpapailaw gamit ang kombensyonal na phototherapy na fluorescent hanggang sa maging normal ang lebel ng kanyang bilirubin. Sa aking pagpayag na ilahok ang aking anak sa pag-aaral na ito ay nauunawaan ko rin ang aking papel na gagampanan na sagutan ang monitoring sheet kung saan aking itatala ang oras kung kalian magagambala ang pagpapailaw sa kanya sa ilalim ng LED phototherapy tulad ng oras ng pagpapadede, pagpapaligo at pagkuha sa kanya upang kuhanan ng dugo.

Naunawaan ko na posibleng magkaroon ng mga di inaasahang reaksyon sa bata katulad ng pamumula ng balat o pagkakaroon ng rashes, pagtatae, paglalagnat, pagkatuyot o dehydration. Ganunpaman, naunawaan ko na may masinsinang pagbabantay na gagawin upang maiwasan ang mga posibleng komplikasyon na nabanggit. Kung magkakaroon man ng anumang reaksyon dulot ng pagsusuring ito ay alam ko na bibigyan ito ng kaukulang paglunas ng mga doktor.

Ipinaalam rin sa akin na wala akong makukuhang pinansyal na suporta o pera sa pagsali sa pag-aaral na ito. Tanging libreng serbisyo ng mga doktor at paggamit ng mga pasilidad ng Ospital gaya ng phototherapy unit ang aming makukuha. Naipaliwanag rin sa akin na ang mga impormasyon tungkol sa aking anak ukol sa pagsali n'ya sa pag-aaral na ito ay hindi ibubunyag at mananatiling kumpidensyal na naaayon sa batas, liban sa gumagawa ng pagsusuring ito. Ang mga imposmasyon mula sa pagsusuring ito ay maaring mailathala o maibigay sa ibang taong gumagawa ng pagsusuri, ngunit ang pagkakakilanlan sa akin at sa aking anak ay hindi ibubunyag.

Alam ko na ang pagsali sa pagsusuri ay dapat kusang-loob. Maaari kong bawiin ang pakikilahok sa pagsusuri ng aking anak kahit anong oras at hindi magbabago ang pagbibigay ng lunas sa aking anak. Lahat ng aking katanungan ukol sa pagsusuring ito ay nasagot ng may katiyakan at lubos kong nauunawaan. Binigyan rin ako ng sapat na oras para pag-isipan ang pagbibigay ng aking pahintulot.

Kung ako man ay may katanungan, maaari kong kausapin si Dr. Vanessa Calabia sa telepono bilang: 8826316 loc 258 o cellphone: 09255842009 o si Dr. Viel Bagunu sa telepono bilang: 8826316 loc 258 o cellphone: 09256827835. At kung ako man ay may katanungan sa kung paano nabuo ang LED phototherapy ay maari kong kausapin si Engr. Paul Cabacungan ng Ateneo Innovation Center sa telepono bilang: 9272147 o cellphone: 09183860216.

Sa aking paglagda ay ipinapahiwatig nitong nabasa at naunawaan ko ang mga nakatala at kusang-loob na pinahihintulutan ko ang aking anak na lumahok sa naturang pag-aaral.

______________________________        ______________

Pangalan ng pasyente          Petsa

______________________________        ______________

Pangalan at lagda ng magulang          Petsa

______________________________        ______________

Pangalan at lagda ng saksi          Petsa

________________________________        ______________

Pangalan at lagda ng kumukuha ng pagsang-ayon          Petsa

Appedix B.

## NOMOGRAM

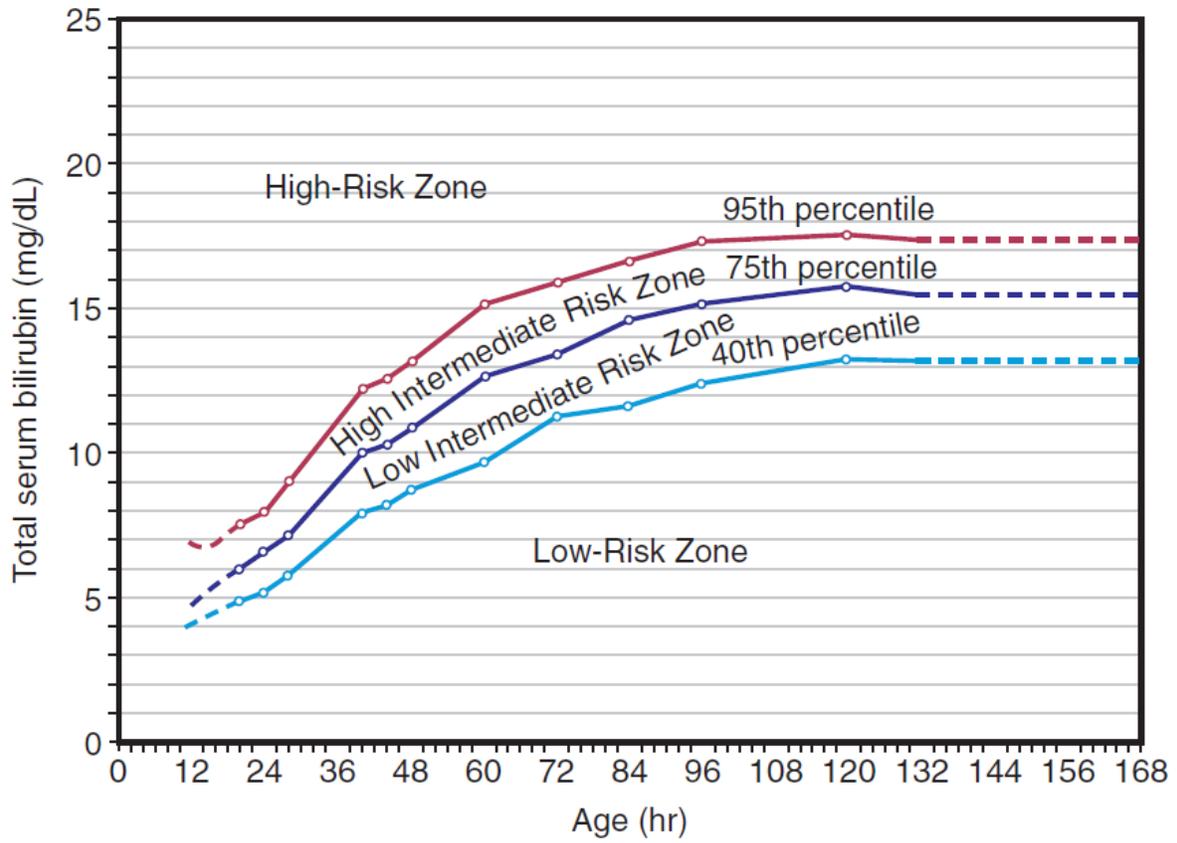

**Figure 1: NOMOGRAM** Risk designation of term and near-term well newborns based on their hour-specific serum bilirubin values. The high-risk zone is subdivided by the 95th percentile track. The intermediate -risk zone is subdivided into upper and lower risk zones by the 75th percentile track. The low-risk zone has been electively and statistically defined by the 40th percentile track. (From Bhutani VK, Johnson L, Sivieri EM: Predictive ability of a predischarge hour-specific serum bilirubin for subsequent significant hyperbilirubinemia in healthy term and near-term newborns, Pediatrics 103:6–14, 1999.) SOURCE: Nelson Textbook of Pediatrics 19th edition

Appendix C.

# Case report form

Name of Participant: _____________________ Patient number: ________

Age /sex:___________ Date/Time of birth:_____________________

Patient's blood type: ________ Mother's blood type: _________

Baseline TSB: _____________ Feeding: ____________

Phototherapy:

Date/time started: _____________ Date/time ended: _________________

| Phototherapy (Hours) | TSB (mg/dL) | Risk Zone (based on hour-specific normogram) | Complications | | | |
|---|---|---|---|---|---|---|
| | | | Rash | Blister | Watery stool | Others |
| 24h | | | | | | |
| 48h | | | | | | |

Appendix D.

## **Mother Data Collection Form**

Name of Patient: _______________________________________________

|  | Breastfeeding | Cleaning | Turning | Blood extraction |
|---|---|---|---|---|
| Date/time |  |  |  |  |
|  |  |  |  |  |
|  |  |  |  |  |
|  |  |  |  |  |

Appendix E.

## Patient Monitoring Form

Name of Patient: ______________________________________________

| Date/time | Weight | Temperature | Urine output |
|-----------|--------|-------------|--------------|
|           |        |             |              |
|           |        |             |              |
|           |        |             |              |

Appendix F.

# PHOTOTHERAPY LIGHT FOR JAUNDICE TREATMENT

Vanessa Marie V. Calabia MD*, Gregory L. Tangonan, Ph.D.**., Jeremy E. De Guzman MD*, MBA, ,  Paul M. Cabacungan, MS** Ivan B. Culaba, MS***

*Ospital ng Makati

**Ateneo Innovation Center, Ateneo de Manila University

***Department of Physics, Ateneo de Manila University

Neonatal jaundice occurs in 25% to 50% of term newborns, and in a larger proportion of preterm newborns, in the first two weeks of life.  It is a benign transient physiological event in the majority of newborns but can cause irreversible brain damage and kernicterus in some infants if the serum bilirubin levels are very high. Various mechanisms involved in producing this 'physiological' increase in serum total bilirubin include increased production of bilirubin due to lysis of red blood cells, decreased ability of liver cells to clear bilirubin and increased enterohepatic circulation. Any condition that further increases bilirubin production or alters the transport or metabolism of bilirubin increases the severity of the physiological jaundice (2).

Phototherapy is the most frequently used treatment when serum bilirubin levels exceed physiological limits. To initiate phototherapy without delay is the most important intervention for infants with severe hyperbilirubinemia.  Phototherapy converts bilirubin into water soluble photo-products that can bypass the hepatic conjugating system and be excreted without further metabolism. The clinical response to phototherapy depends on the efficacy of the phototherapy device, as well as the infant's rates of bilirubin production and elimination (1).   Phototherapy, despite being the recommended standard of treatment for hyperbilirubinemia in newborns, may not be available in developing countries because the devices and replacement bulbs are often too expensive.

Phototherapy can be delivered using several types of conventional light sources, including daylight, white or blue fluorescent bulbs and filtered halogen bulbs.  The efficacy and ability of these light sources to provide intensive phototherapy varies widely and some may be limited because of the inability to keep them close to the infant due to high heat production and unstable broad wavelength light output. In recent years, a new type of light source, light-emitting diodes (LEDs) have been developed and studied as possible light sources for the phototherapy of neonatal jaundice. LEDs are power efficient, portable devices with low heat production, light in weight and have a longer lifetime (2).  Blue LEDs emit a high intensity narrow band of blue light overlapping the peak spectrum of bilirubin breakdown resulting in potentially shorter treatment times.  These characteristics of LEDs make them an optimal light sources for a phototherapy device (3).

**Specifications of the light source**

    a. Emission spectrum of the LED lamp

A Spectro-Vis Plus (Vernier) was used to determine the spectrum of the LED lamps. Figure 1 below shows the peak intensity at 462.1 nm which falls on the blue region of the visible spectrum of light. The range of the emission spectrum was found to be 462.1-476.4 nm. This range falls within the most effective range (460-490 nm) for phototherapy of jaundice [3]. The instrument used in determining the emission spectrum of the lamps is shown in Figure 2.

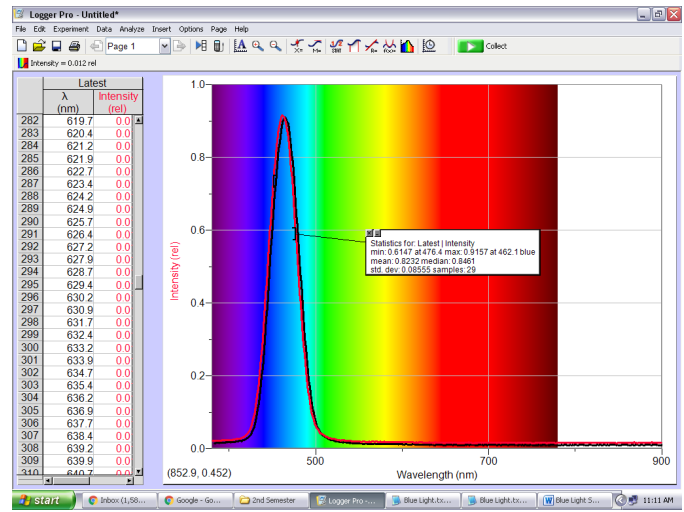

Figure 1. The emission spectrum of the light sour

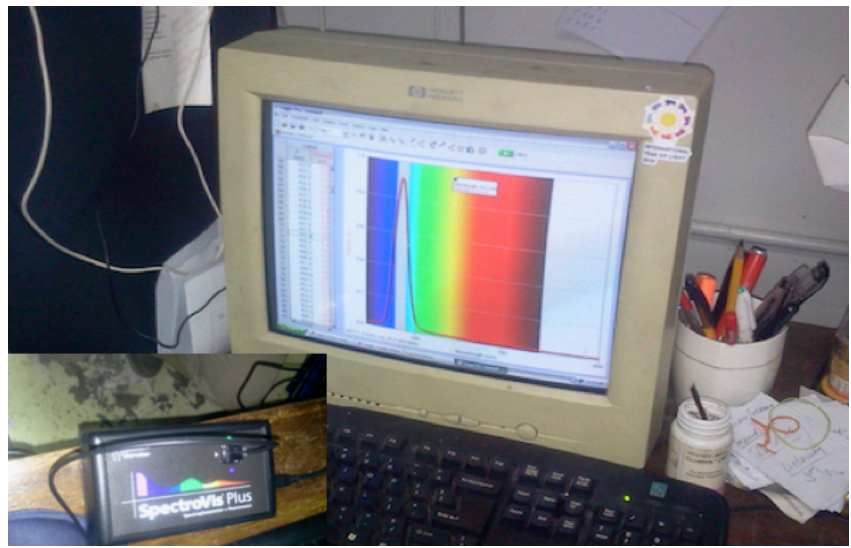

Figure 2. The Spectro-Vis Plus used in determining the emission spectrum

b. Irradiance

The intensity of the LED was 12W/m² measured using a light intensity meter.

The calculated value of the irradiance was 84 µW/cm²/nm. This value is almost three times higher than the minimum required irradiance (30 at µW/cm²/nm) at 1-m distance. Note that protocol for clinical trial is only 30 cm and therefore the irradiance of this light source is sufficient for phototherapy. The calculation of the irradiance is shown below. Figure 3 shows the measurement of the intensity.

Irradiance = measured intensity (microwatt/ square cm) / range of wavelength

Irradiance = {(12 W/ m²) * (10⁶ µW / watt) * [(1 m²/10⁴cm²)]}

/ {476.4 nm– 462.1 nm}

=84 µW /cm²/nm

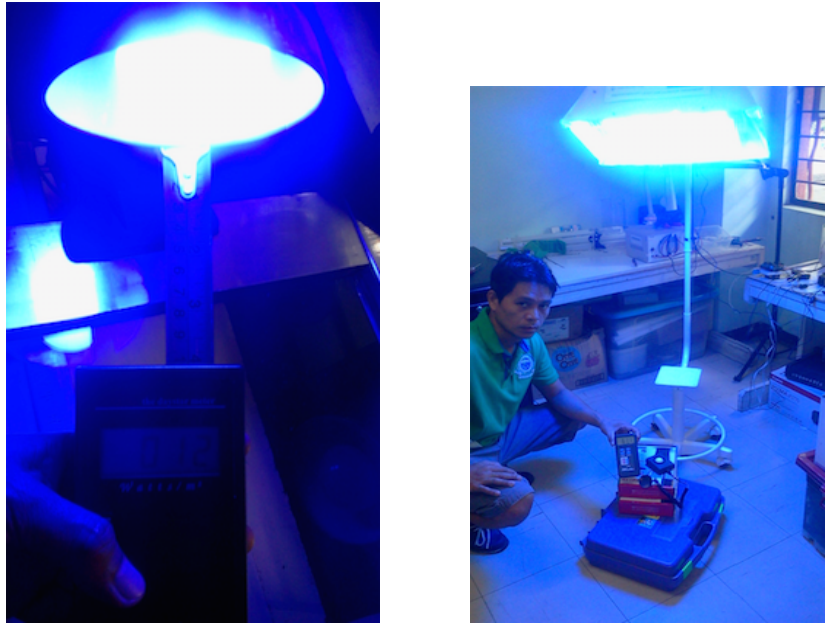

Figure 3. Measurement of the intensity of the LED lamp

c. Electrical power requirement

Each lamp is operated at 220 V at 3 W. The total power requirement for 20 lamps is 60 W. To maintain low temperature of operation a 10-W computer fan is used for ventilation.

**The Phototherapy Equipment**

Photos of the phototherapy equipment are shown in Figures 4, 5 and 6. There are 20 pieces of blue LED lamps, connected in parallel, for a total power rating of 60 W. The assembly of the lamps is shown on Figure 4.

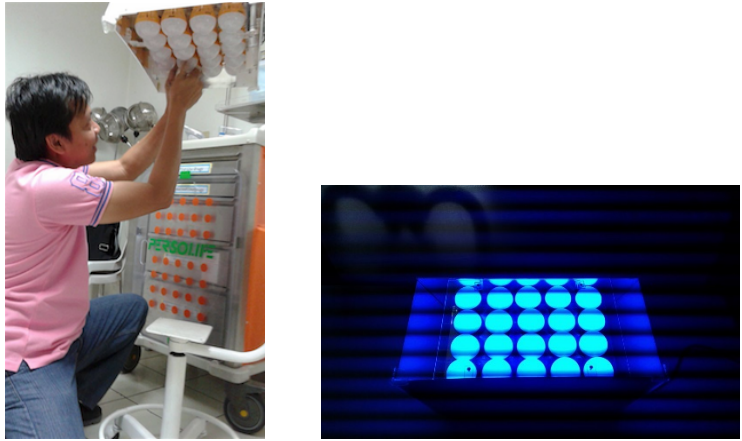

Figure 4. Assembly of the 20 pieces of blue LED lamps.

There is a 2-A fuse for protection due to overcurrent. Ten lamps are turned on/off by a switch and the other ten lamps are turned on/off by a separate switch. A 10-W computer fan was installed for cooling the light casing, ventilation and a safety feature to exhaust any harmful fumes that the circuit might generate. This fan operates even when the lamps are turned off. There is a main power switch which controls all the electrical parts. The top cover has 2-mm diameter perforations, distanced 1 cm apart, which allow warm air to escape from the case. The interior part of the casing is lined with silver foil reflector sheet to maximize the illumination to the target area. There is a clear plastic cover to safely secure the bulbs in place. Hinges were attached at the sides of the light casing for easy maintenance and/or trouble shooting. Figure 5 shows the casing of the phototherapy equipment.

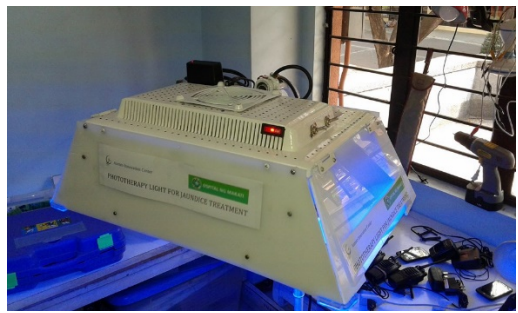

Figure 5. The casing of the light source

The unit is designed in such a way that the whole light casing can be tilted from side to side, the metallic pipe stand height can be adjusted, the whole unit can be rolled through its wheels for mobility/easy handling. Figure 6 shows the complete phototherapy equipment. The total cost for the development of the prototype equipment was approximately PhP 20,000.00

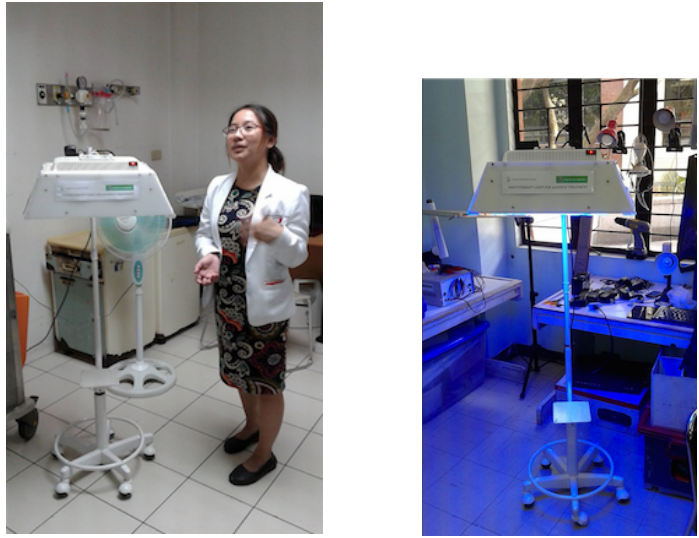

Figure 6. The complete phototherapy equipment

The total power consumption of this prototype is 75 watts which is less than one fifth of the commercial phototherapy light (about 400 watts) that uses fluorescent tubes. This prototype unit can also powered by photovoltaic cells (solar panels). For a 100-watt solar panel at four-hours of sunlight, it can operate for five-hours with all light bulbs continuously switched on. It is designed in such a way that at it can also operate at half power and twice the time.

An internal continuous test-run was conducted with all the lamps turned on for 7-hours, 24-hours, and 72-hours. There was no notable degradation of the quality of the lamp.

References:

1. Managing Newborn Hyperbilirubinemia and Preventing Kernicterus Progeny, Volume XX1X, No 1, June 2013 available in http://www.idph.state.ia.us/hpcdp/common/pdf/perinatal_newsletters/progeny_june2013.pdf

2. Kumar P, Chawla D, Deorari A. Light-emitting diode phototherapy for unconjugated hyperbilirubinaemia in neonates. Cochrane Database of Systematic Reviews 2011, Issue 12. Art. No.: CD007969. DOI: 0.1002/14651858.CD007969.pub2

3. Chang YS, et. al. In vitro and in vivo Efficacy of New Blue Light Emitting Diode Phototherapy Compared to Conventional Halogen Quartz Phototherapy for Neonatal Jaundice. J Korean Med Sci 2005; 20: 61-4